\documentclass[twocolumn]{aastex701}
\usepackage{hyperref,natbib}

\newcommand{\grb}{GRB\,260127A}
\newcommand{\swift}{\textit{Swift}}

\begin{document}

\received{March 10, 2026}
\revised{March 26, 2026}
\accepted{April 3, 2026}
\submitjournal{ApJL}

\title{Rapid-response 1.3~mm Observations of \grb\ with the Submillimeter Array}
\shorttitle{Rapid Millimeter Observations of \grb}
\shortauthors{Keating et al.}

\newcommand{\CfA}{\affiliation{Center for Astrophysics \textbar{} Harvard \& Smithsonian, 60 Garden Street, Cambridge, MA 02138-1516, USA}}

\correspondingauthor{Garrett K.~Keating}
\email{garrett.keating@cfa.harvard.edu}

\author[0000-0002-3490-146X]{Garrett K.~Keating}
\email{garrett.keating@cfa.harvard.edu}
\CfA

\author[0000-0003-1792-2338]{Tanmoy Laskar}
\email{tanmoy.laskar@utah.edu}
\affiliation{Department of Physics \& Astronomy, University of Utah, Salt Lake City, UT 84112, USA}

\author[0000-0002-9017-3567]{Anna Y. Q.~Ho}
\email{ayh24@cornell.edu}
\affiliation{Department of Astronomy, Cornell University, Ithaca, NY 14853, USA}

\author[0000-0003-0526-2248]{Peter K.~Blanchard}
\email{peter.blanchard@cfa.harvard.edu}
\CfA

\author[0000-0002-8297-2473]{Kate D.~Alexander}
\email{kdalexander@arizona.edu}
\affiliation{Department of Astronomy/Steward Observatory, 933 North Cherry Avenue, Rm. N204, Tucson, AZ 85721-0065, USA}

\author[0000-0002-9392-9681]{Edo~Berger}
\email{eberger@cfa.harvard.edu}
\CfA

\author[0000-0003-0685-3621]{Mark~Gurwell}
\email{mgurwell@cfa.harvard.edu}
\CfA

\author[0000-0003-0307-9984]{Tarraneh Eftekhari}
\email{teftekhari@northwestern.edu}
\affiliation{Center for Interdisciplinary Exploration and Research in Astronomy, Northwestern University, 1800 Sherman Avenue, Evanston, IL 60201, USA}

\author[0009-0009-1572-6254]{Chloe T.~Xu}
\email{chloex@mit.edu}
\affiliation{Department of Earth, Atmospheric, and Planetary Sciences, Massachusetts Institute of Technology, Cambridge, MA 02139, USA}

\author[0000-0002-4248-5443]{Joshua Bennett Lovell}
\email{joshualovellastro@gmail.com}
\CfA

\author[0000-0002-1407-7944]{Ramprasad Rao}
\email{rrao@cfa.harvard.edu}
\CfA

\author[0000-0003-3734-3587]{Peter K.~ G.~Williams}
\email{pwilliams@cfa.harvard.edu}
\CfA

\begin{abstract} 
We present the results from rapid-response 1.3~mm observations of \grb\ using the Submillimeter Array (SMA). SMA arrived on-source 12.6 minutes after the initial detection by the Neil Gehrels Swift Observatory, representing the earliest millimeter/submillimeter observations of a GRB to date. From these observations, we find a source with flux density $6.9\pm1.7$ mJy, consistent with the X-ray afterglow position but slightly offset from the optical afterglow position (2.7$^{\prime\prime}$ offset, with the SMA detection having a 90\% confidence radial position uncertainty of 0.9$^{\prime\prime}$). Subsequent observations 1.9 days later show no sources of emission, with a $3\sigma$ upper limit of 0.70 mJy. If the SMA detection is associated with \grb, we infer that the 1.3~mm light curve for \grb\ declined at least as fast as $t^{-0.5}$, suggesting that peak brightness of the event at this wavelength was reached in under a day. We discuss how these findings may be consistent with both forward shock and reverse shock afterglow scenarios, and implications for future millimeter/submillimeter observations of GRBs on these timescales.
\end{abstract}


\section{Introduction}\label{sec:intro}
Gamma-ray bursts (GRBs) are energetic flashes of $\gamma$-rays lasting from fractions of a second to minutes \citep{kso73}. While they are known to be powered by transient jets launched in the core collapse of massive stars or merging compact objects, the true nature (e.g., structure, composition, and launching mechanism) of these jets remains uncertain \citep{rm94}. The interaction of the jet with the environment produces shocks propagating in the ambient medium -- the forward shock (FS) -- and a reverse shock (RS) propagating back into the ejecta \citep{sar97,kps99}. Both shocks are expected to accelerate electrons to relativistic energies, producing detectable synchrotron radiation \citep{spn98}. Since the FS emission is sensitive only to the explosion energy and agnostic to other properties of the jet, RS radiation remains a key technique to study the composition, magnetization, and Lorentz factor of GRB jets \citep{kz03,mkp06}. RS emission is expected to manifest as an optical flash and a radio flare \citep{abb+99,mr99,fdhl02,wei03,2026MNRAS.tmp..790S}. Searches for RS emission at optical wavelengths have yielded a handful of candidates \citep{2008Natur.455..183R, Vestrand2014, MartinCarrillo2014, Becerra2019, Zhang2022, Jayaraman2024}. RS emission has also been detected at lower frequencies\footnote{The RS peak frequency is expected to be lower than that of the FS by a factor of $\sim\Gamma^2$} \citep{mmk+08}, particularly the radio, with inferred jet Lorentz factors $\approx10^2$ commensurate with expectations from prompt $\gamma$-ray properties \citep{lbz+13,pcc+14}. At millimeter (mm) wavelengths, fast-fading, RS-like emission has been seen in a few cases, allowing for stronger constraints on the RS peak (and hence on the RS evolution and the jet parameters; \citealt{lab+16,lves+19,lag+19}); however, the short-lived nature of the RS, combined with strong proposal pressure on facilities sufficiently sensitive to capture this emission (e.g., ALMA) for non-ToO science have hobbled studies at these wavelengths.

We have recently commissioned a new rapid-response mode with the Submillimeter Array (SMA). This new mode allows for triggered observations to be interleaved with more ``standard'' observations in a semi-automated fashion, enabling the array to make milliJansky-level measurements in a matter of minutes, with minimal impact on other science programming. We here report on the first scientific results using this new capability, in which SMA successfully conducted rapid-response observations of \grb, arriving on source minutes after initial detection by the Neil Gehrels Swift Observatory (\textit{Swift}; \citealt{Gehrels2004}).

We have structured this manuscript in the following way. We describe the observations and data collected in Section~\ref{sec:obs}. We discuss the results of our data analysis, results, and data verification tests in Section~\ref{sec:analysis}. These results are discussed in further detail in Section~\ref{sec:disc}, with conclusions presented in Section~\ref{sec:conclusion}. Unless otherwise noted, throughout this letter we report uncertainties in terms of the standard deviation assuming Gaussian statistics (i.e., $\pm1\sigma$; 68.3\% uncertainty). For reference, we note that \grb\ was initially detected by \swift\ at 2026 January 27 at 17:50:34 UT \citep[denoted as $T_0$ hereafter;][]{2026GCN.43529....1K}.
\begin{figure*}[!bhtp]
    \centering
    \includegraphics[width=0.9\linewidth]{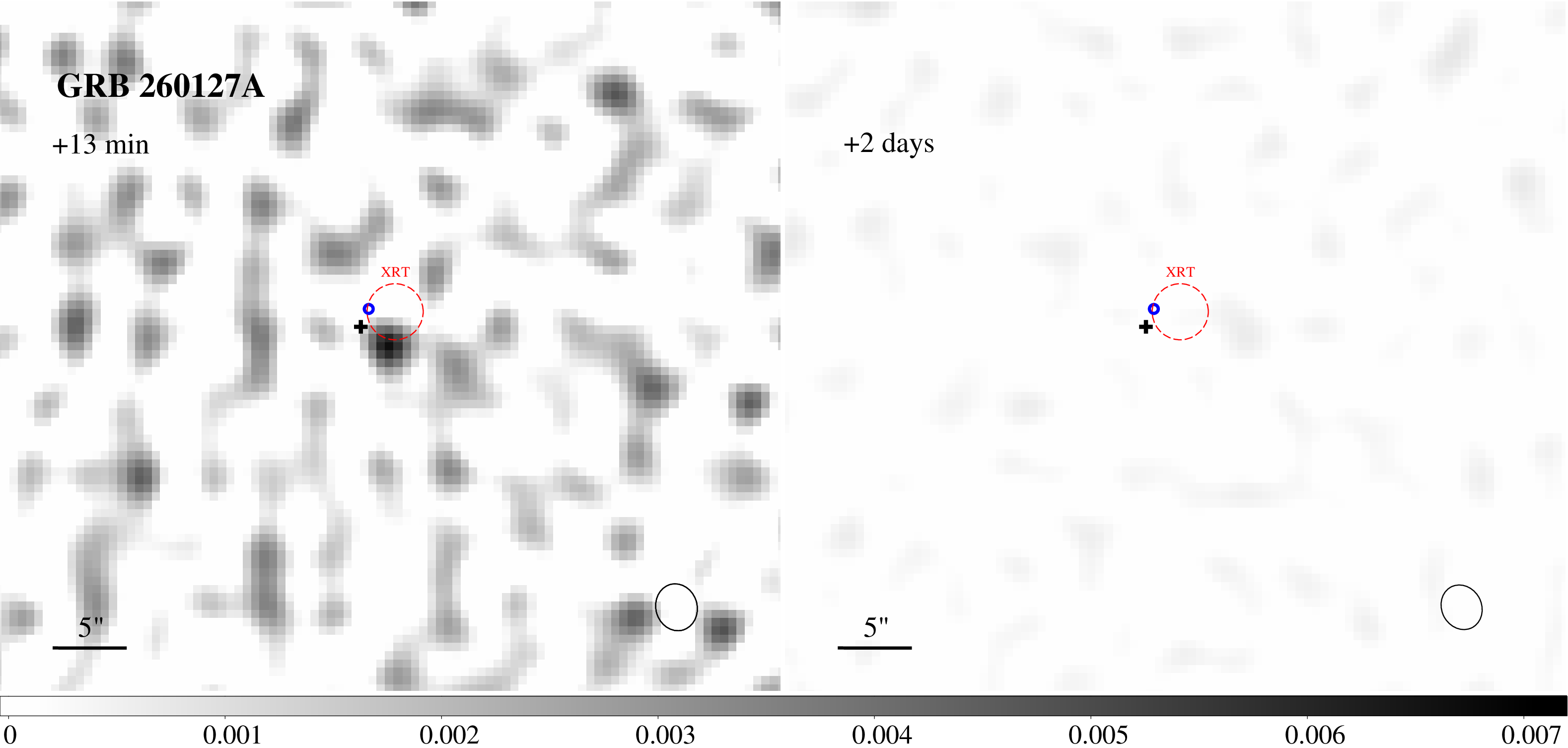}
    \caption{
    \emph{Left}: Image from SMA observations of \grb\ on January 27 (10-min observation). A single point source-like feature at high significance ($4.2\sigma$) is seen, near the the reported X-ray and optical afterglow positions, as measured by \swift/XRT (dashed red circle; 90\% error radius) and LCO (solid blue circle). There is a faint ($g\sim r \sim 23\,$mag) cataloged source in the Legacy Survey \citep{Dey2019}, with an extended morphology, 1.3$^{\prime\prime}$ from the optical position and 2.2$^{\prime\prime}$ from the SMA position (marked with a plus sign). The synthesized beam is shown in the lower right, with size $3.2^{\prime\prime}\times 2.8^{\prime\prime}$ (PA: $14.8^{\circ}$).
    \emph{Right}: Image from the January 29 SMA observations (4-hour observation). No significant source is seen around the position of \grb, nor are any significant sources of emission seen within the field of view of the telescope. Note that the noise level in this image is much lower than that from the January 27 data due to increased observing time and better weather conditions. The synthesized beam is similar to the first observation ($3.1^{\prime\prime} \times 2.7^{\prime\prime}$; PA: $27.1^{\circ}$).}
    \label{fig:images}
\end{figure*}
\section{Observations}\label{sec:obs}
On 2026 January 27, SMA was running ``standard'' observations in the vicinity of MWC\,349A, with the array tuned to an local oscillator (LO) frequency of 225.538 GHz and configured for polarimetric observations, with 8 elements in the compact configuration, with typical baseline lengths of 10--70 m. At 17:54 UT ($\sim T_{0}+200\,\textrm{s}$), SMA observations of \grb\ were triggered following the initial detection from \swift\ Burst Alert Telescope (BAT; \citealt{Barthelmy2005}), and subsequent detection \citep{2026GCN.43529....1K} by the \swift\ X-Ray Telescope (XRT; \citealt{swift_xrt}).

Subsequent analysis of \swift/XRT data yielded a final position for the X-ray afterglow of $(\alpha, \delta)_{\rm J2000}=$ $15^{\rm h}27^{\rm m}31.02^{\rm s}$ $+06^{\circ}45^{\prime}45.3^{\prime\prime}$ \citep{2026GCN.43539....1D,2007A&A...476.1401G,2009MNRAS.397.1177E}. The 90\% error radius on the XRT position is 1.9$^{\prime\prime}$. Observations with the Las Cumbres Observatory (LCO) 1-m telescope detected the optical afterglow, with a best-determined position $(\alpha, \delta)_{\rm J2000}=$ $15^{\rm h}27^{\rm m}31.14^{\rm s}$ $+06^{\circ}45^{\prime}45.49^{\prime\prime}$ \citep{2026GCN.43530....1S}. The reported 1-$\sigma$ uncertainty on the optical position is 0.3$^{\prime\prime}$.

SMA observational planning was performed by a prototype piece of software known as ASTROLABE (Automated Software for Triggering Rapid Observations with Legions of Astrophysically Bright Events), which identified 1550+054 as an appropriate nearby complex gain calibrator, with bandpass and flux density calibration measurements -- using 3C\,279 and MWC\,349A, respectively -- leveraged from the interrupted standard observations. Triggered observations began at 17:58 UTC ($\sim T_0+500\,\textrm{s}$) after bringing previously executing observations to a phased halt. SMA initially slewed to the gain calibrator, and subsequently arrived on \grb\ at 18:03:22 UT ($T_0+758\,\textrm{s}$). To minimize impact on concurrent programs, triggered observations were limited to a single dwell on-source, which concluded at 18:13:12 UT ($T_0+1348\,\textrm{s}$), for a total exposure time of 590 s. Following a final observation of 1550+054, the previously running observations were resumed.

Based on preliminary findings (discussed further in Section~\ref{ssec:results}), follow-up observations -- scheduled under the SMA program {\it POETS} (Pursuit of Extragalactic Transients with the SMA; 2022B-S046; PI: Berger) -- were conducted on 2026 January 29, between 14:12 and 20:17 UT. An LO tuning of 225.5 GHz was used in dual-polarization mode, with 8-elements operating. One of the pair of receivers on Antenna 4 experienced issues during bandpass and flux calibrator observations, and was subsequently flagged during data processing. On this date, \grb\ was observed for a total of 3.2 hours. 1550+054 and 1513+002 were used as complex gain calibrators, 3C\,279 and 3C\,454.3 were used as bandpass calibrators, and MWC\,349A was used as the flux density calibrator.

\section{Data Analysis}\label{sec:analysis}
Data for both observations were processed via the SMA Calibrator Observations for Measuring the Performance of Array Sensitivity and Stability (COMPASS) pipeline (Keating et al., \textit{in prep.}; \citealt{2015ApJ...814..140K}), which automatically flags outliers in amplitude (in both spectral and temporal domains), flags baselines that show little or no coherence on calibrator targets, and derives antenna-based bandpass and gains solutions. Flux calibration was performed with the COMPASS fiducial model\footnote{Based in part on previously published analysis of SMA data \citep{2023ApJ...953L...6P} as well as ongoing observations of MWC\,349A.} for MWC\,349A ($S_{\nu}=2.0\cdot(\nu_{\rm obs} / {\rm 225\ GHz})^{0.6}\,\textrm{Jy}$). The data were imaged using natural weighting, and deconvolution was performed via the CLEAN algorithm \citep{1974A&AS...15..417H}.

\subsection{Results}\label{ssec:results}
The image resulting from the January 27 data is shown in the left panel of Figure~\ref{fig:images}. In the image we detect a point source in this first epoch at position $(\alpha,\delta)_{\rm J2000}=$ $15^{\rm h}27^{\rm m}31.05^{\rm s}$, $+06^{\circ}45^{\prime}43.1^{\prime\prime}$, with a fitted peak flux density of $6.7\pm1.6$ mJy. Fitting the visibilities from January 27 with a spectrally-flat point source model yields a consistent result, with flux density of $6.9\pm1.7$ mJy, and a 90\% radial position uncertainty of 0.9$^{\prime\prime}$. The SMA position is offset from the X-ray afterglow position by $2.3^{\prime\prime}$, and from the optical afterglow position by $2.7^{\prime\prime}$ -- the measured flux at these positions is consistent with a null detection, with a $3\sigma$ upper limit of 4.8 mJy.  These results were verified with FLARE (Fitting Library for Astrophysical Radio Emission; Xu et al. \emph{in prep.}), which provides a Bayesian fitting framework for automated source finding via peak detection in the image domain and model fitting in the visibility domain.

An image was also produced from the January 29 data, which is shown in the right panel of Figure~\ref{fig:images}. No sources of emission at either the XRT, optically, or mm-defined positions, with a $3\sigma$ upper limit of 0.70 mJy, and no sources within an area twice the field of view down to a limit of $S_{\rm 225 GHz}\leq0.9$ mJy.
\subsection{Data Validation}\label{ssec:validate}
\begin{figure}[t]
    \centering
    \includegraphics[width=1.1\linewidth]{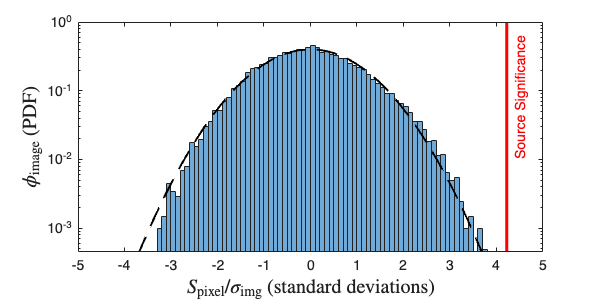}
    \caption{A histogram of the pixel fluxes ($S_{\rm pixel}$) from imaging the January 27 data after subtraction of the fitted point-source model from the visibilities. The total area imaged in this analysis is approximately $3\times$ the area of the primary beam (i.e., measuring $81^{\prime\prime}$ on a side). The distribution of pixel fluxes are in good agreement with theoretical expectations for Gaussian noise (black dashed line), with the source significance (red line) modestly separated from the distribution of noise-like values in the image.}
    \label{fig:histogram}
\end{figure}
As these observations were the first of its kind, demonstrating a new novel capability for SMA, additional checks were performed on the data in order to verify the results. Data from the January 27 observation were split by time, polarization, and sideband, with the result being consistent between the split samples. We also imaged the data after dropping each antenna, and the resulting fluxes are consistent with that is shown in Figure~\ref{fig:images}. Further checks were limited by signal-to-noise, though checks of visibility amplitudes revealed no obvious outliers. Histogram analysis of the resulting image, shown in Figure~\ref{fig:histogram}, indicates Gaussian-like behavior of the noise in the image domain. Imaging of a subset of the January 29 data around the same hour angle range as the January 27 observations also show no indication of a source near the afterglow position or elsewhere in the field. As such, there are no indications that the source detected on January 27 is spurious.
\subsection{Astrometry Checks}\label{ssec:astrometry}
Due to the apparent difference between the SMA-detected mm source and the optical afterglow positions of \grb, additional astrometry checks were performed on the data presented here. We note that positional errors for the SMA measurements are derived from fitting uncertainties, and do not account for potential systematics that may be present.

Checks on calibrator data from both the January 27 and 29 observations show no indication of issues at the level of $\leq0.1^{\prime\prime}$. The positions of gain calibrators used here are consistent with that of other observatories\footnote{\url{https://science.nrao.edu/facilities/vla/observing/callist}}. Recorded telemetry for the January 27 observations is consistent with expectations, though we note that these checks -- while extensive -- should not be considered exhaustive due to the complex nature of the control systems for SMA. Real time astrometry for SMA is calculated using \emph{SuperNOVAS}\footnote{\url{https://sigmyne.github.io/SuperNOVAS/}}, which has been checked against packages such as astropy at the sub-milliarcsecond level. Previous studies of transients (using the same analysis tools and similar workflow) have shown typical astrometric agreement to $\ll 1^{\prime\prime}$, with respect to the transient or other known sources of emission within the field \citep[e.g.,][]{2023ApJ...951L..31B,lam+23}, even when offset from the phase center of the array \citep[e.g.,][]{2025ATel17290....1H}.

However, we note that these example observations extended over multiple hours, and that the precision of astrometry in ``snapshot'' observations is less explored on SMA. Moreover, ambient weather conditions on Maunakea were shifting around 18 UTC on January 27, with an unusually sharp $\approx20\%$ drop in relative humidity measured around the 15-minute window of the observations of \grb. Snapshot imaging of another source observed during this period (immediately prior to triggered observations of \grb) do show a $1.4^{\prime\prime}$ shift in position in a direction consistent with the mm-optical offset. While it is unclear if changing weather conditions could have introduced an offset of order a resolution element of the interferometer, it is conceivable that these conditions (along with the snapshot nature of the observations) could give rise to previously unexplored systematic errors, which would lead us to underestimate the positional uncertainty in the SMA measurement. As there were no sources of emission in the field beyond \grb\ to register SMA measurements against, the addition of a second gain calibrator or an astrometry ``check source'' would have benefited the January 27 observations, and could prove useful in future use of the SMA rapid response mode.\section{Discussion}\label{sec:disc}
\begin{figure}[!t]
    \centering
    \includegraphics[width=0.9\linewidth]{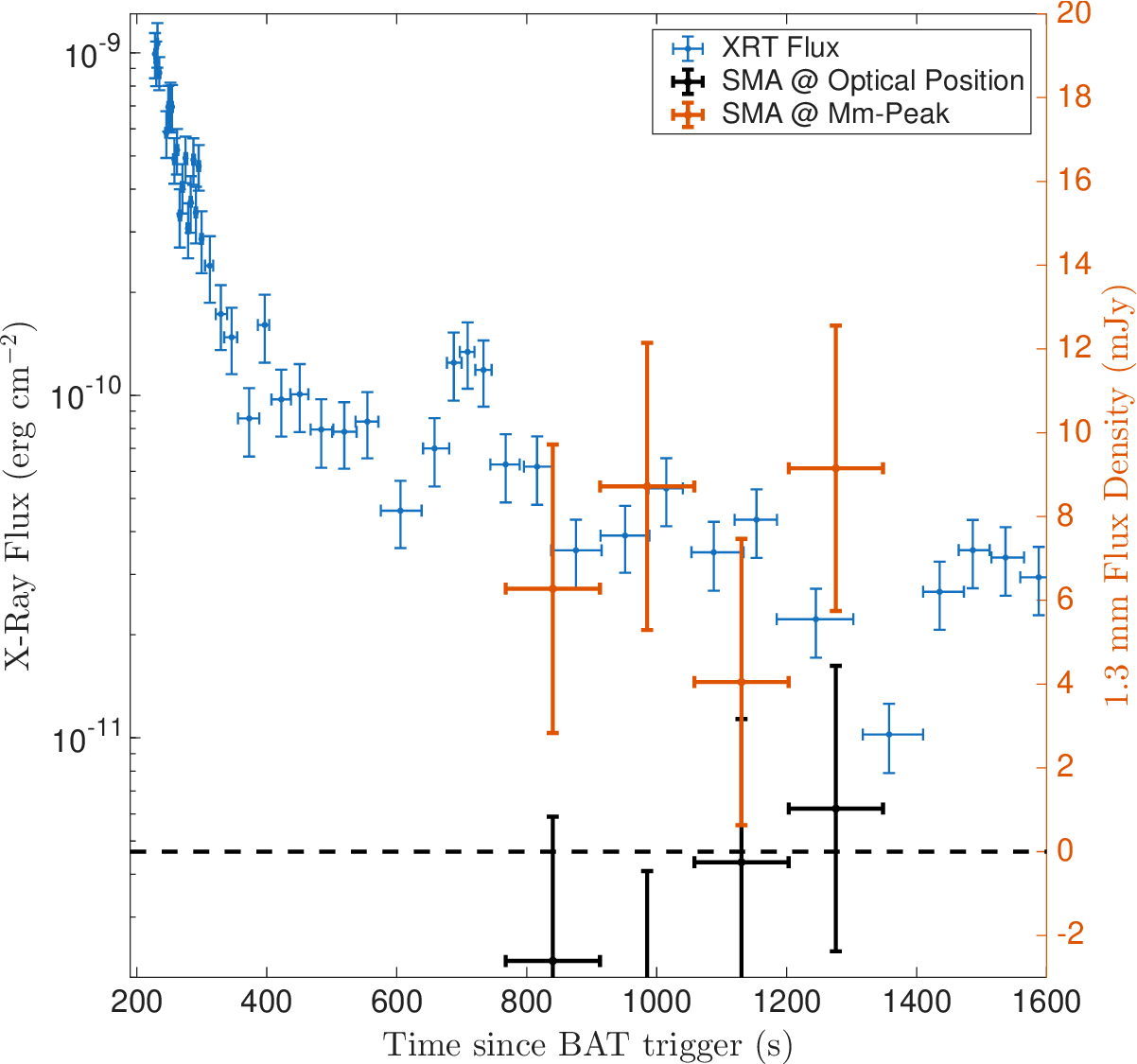}    
    \\
    \
    \\
    \includegraphics[width=0.9\linewidth]{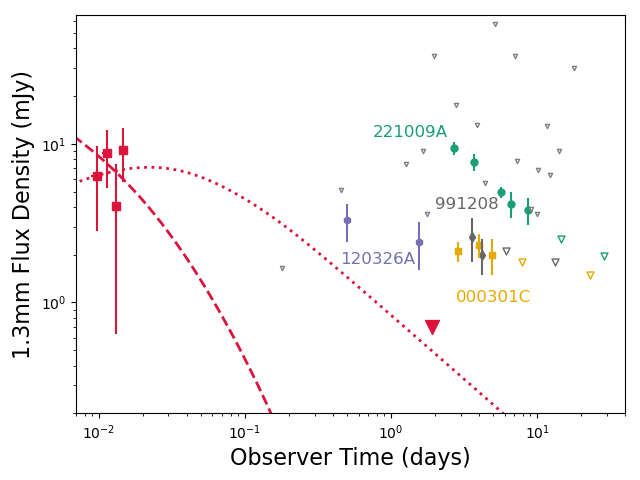}
    \caption{\emph{Top}: Light curve data of \grb\ as measured by \swift\ XRT \citep[blue;][]{2007A&A...469..379E,2009MNRAS.397.1177E}, and by SMA at both the optical afterglow position (black) and the mm-bright position (orange). \emph{Bottom}: SMA 225~GHz observations of \grb\ (red points) along with fiducial RS (dashed) and FS (dotted) models, compared with a sample of ($235\pm50$)~GHz light curves, with detections (colored points) highlighted (from \citealt{bsf+00,gbb+00,duplm+12,lbm+15,lam+23,brf+23}). Triangles indicate upper limits.}
    \label{fig:data_products}
\end{figure}
These SMA observations represent the earliest mm/sub-mm observations of a GRB to date (Figure~\ref{fig:data_products}). Direct comparison of light curve behavior to that of other events is difficult, both due to the limited number of previous observations at 1.3~mm (number of events and cadence at which those events have been measured). Although additional events have been studied at 3~mm and 850 {\textmu}m, including some within hours of the burst \citep[e.g.,][]{2011GCN.11580....1Z,2012GCN.13233....1S}, the prior lack of minutes-timescale mm/sub-mm response makes comparison challenging.

Nevertheless, the SMA non-detection at 1.92~days in this case implies that the 1.3~mm light curve in this case declines at least as fast as $\sim t^{-0.5}$, suggesting that the 1.3~mm light curve peaked at $\lesssim1$~day in this case.  The inferred minimum decay rate is compatible with mm/sub-mm light curves of prior events observed at later times. For instance, the flux density of GRB\,221009A at 1.3~mm decayed as roughly $S_{\nu} \propto t^{-0.8}$, although on much longer timescales of $3$--30~days \citep{lam+23}. In their analysis of reverse shock emission from GRB\,221009A, \cite{brf+23} found that 1.3~mm peak would have been observed at $\sim0.4$\,h after the burst -- coincidentally matching the timescales probed here.

In Fig.~\ref{fig:data_products}, we also show a fiducial RS model with initial Lorentz factor, $\Gamma_0\approx150$ and a nominal FS model to guide the eye. We find that both such models can explain the data, but caution that a complete explanation requires multi-wavelength modeling beyond the scope of this work. In particular, we note that the \textit{Swift}/XRT light curve exhibits flaring activity from $\approx3$~min until the first orbital gap at $\approx30$~min, along with a strong re-brightening at $\gtrsim20$~min, possibly indicative of complex outflow structure or viewing geometry, meriting more detailed analysis. Earlier observations along with longer monitoring past the first hour would help distinguish between RS and FS-like models of GRB mm/sub-mm afterglows in the future. 

\section{Conclusions}\label{sec:conclusion}
In this paper we have presented the results of rapid-response observations of \grb\ with the Submillimeter Array at 1.3~mm. These observations, which began 12.6 minutes after initial detection by \swift, yielded a detection of a source near the optical and X-ray afterglow positions, with flux density $6.9\pm1.7$ mJy. This source was not seen in subsequent SMA observations conducted two days later. We find that the 1.3~mm decay rate is at least $\propto t^{-0.5}$, consistent with previously studied events, and discuss that the observed light curve is consistent with models of both FS and RS afterglows. While tantalizing, the results presented here motivate further observational follow-up, both of a wider swath of GRB events as well as more extensive observations of individual events.

These observations were carried out as part of the first successful demonstration of rapid-response capabilities of SMA. While the 13-minute response time represents a major milestone, further improvements to the system are expected to yield responses as quick as 2--3 minutes, with the capability of grouping observations of multiple objects in series for high throughput. These improvements -- which primarily reflect improved automation alongside enhanced science programming policies for the telescope -- stand to benefit studies of many extragalactic transient classes that are bright at mm wavelengths \citep{ebm+22}, including GRBs, where additional follow-up on minutes-to-hours timescales is poised to provide unique insights into such events. Such observations could also be valuable for other classes of extragalactic transients such as X-ray flashes (e.g., \citealt{Sakamoto2005}) and young supernovae (e.g., \citealt{Maeda2021}). 

This new observational mode may also prove invaluable for studies of lower-energy transients, such as stellar flares \citep[e.g.,][]{2021ApJ...911L..25M,2024ApJ...962L..12L}, classical novae \citep[e.g.,][]{2025ATel17290....1H}, and luminous red novae \citep[e.g.,][]{2018A&A...617A.129K}. Moreover, studies of time-variable objects -- such as blazars, active galactic nuclei, and supermassive black holes \citep[e.g.,][]{2025A&A...695A.217M,2026ApJ...997..282M},  as well as black hole X-ray binaries \citep[e.g.,][]{2024NatAs...8.1031V} -- also stand to benefit from these newly demonstrated capabilities.
\begin{acknowledgements}
We thank the referee for their useful and thoughtful feedback, which improved the quality of the manuscript. We thank the several members of SMA staff who provided useful and thoughtful feedback over the course of this project, including K.~Lagua\~{n}a, A.~Mills, P.~Grimes, S.~Paine, S.~Schimpf, R.~Smith, N.~Smith, H.~Thomas, D.~Wilner, and Q.~Zhang. We also thank S.-F.~Yen for their prompt support in carrying out the rapid-response observations presented here. GKK also wishes to thank F. Fornasini and R. Srinivasan for their support, which allowed this project to come to fruition, and J. Weintroub, for his inspiration \`{a} la Timbuk 3. JBL acknowledges the Smithsonian Institute for funding via a CfA J.C. Ryan Fellowship.

The Submillimeter Array is a joint project between the Smithsonian Astrophysical Observatory and the Academia Sinica Institute of Astronomy and Astrophysics and is funded by the Smithsonian Institution and the Academia Sinica. We recognize that Maunakea is a culturally important site for the indigenous Hawaiian people; we are privileged to study the cosmos from its summit.
\end{acknowledgements}

\facilities{SMA, Swift (XRT)}

\software{Astropy \citep{astropy:2013, astropy:2018, astropy:2022}, CASA \citep{CASA}, pyuvdata \citep{2017JOSS....2..140H,2025JOSS...10.7482K}.
}

\bibliography{bibliography,grb_alpha}{}
\bibliographystyle{aasjournalv7}

\end{document}